\pdfoutput=1

\documentclass[wcp]{jmlr}

\usepackage{longtable}

\usepackage{booktabs}
\usepackage{color}
\usepackage[load-configurations=version-1]{siunitx} 
\usepackage{bm}

\jmlrvolume{1}
\jmlryear{2016}
\jmlrworkshop{NIPS 2016, Time Series Workshop}

\title[SeDMiD for Confusion Detection]{SeDMiD for Confusion Detection: Uncovering Mind State from Time Series Brain Wave Data}

  \author{\Name{Jingkang Yang} \Email{jk.yang1995@gmail.com}\\
  \addr   International School\\ Beijing University of Posts and Telecommunications\\ Beijing, 100876, China
  \AND
  \Name{Haohan Wang} \Email{haohanw@cs.cmu.edu}\\
  \addr Language Technologies Institute\\ School of Computer Science\\ Carnegie Mellon University \\Pittsburgh, PA, 15206, the USA
  \AND
  \Name{Jun Zhu} \Email{dcszj@mail.tsinghua.edu.cn  }\\
  \addr State Key Laboratory of Intelligent Technology and Systems\\
Tsinghua National Laboratory for Information Science and Technology\\
Tsinghua University\\
Beijing, 100084, China
  \AND
  \Name{Eric P. Xing} \Email{epxing@cs.cmu.edu}\\
  \addr Machine Learning Department\\ 
  School of Computer Science\\ 
  Carnegie Mellon University \\
  Pittsburgh, PA, 15206, the USA
 }

\editor{List of editors' names}

\begin{document}

\maketitle

\begin{abstract}
Understanding how brain functions has been an intriguing topic for years. With the recent progress on collecting massive data and developing advanced technology,  people have become interested in addressing the challenge of decoding brain wave data into meaningful mind states, with many machine learning models and algorithms being revisited and developed, especially the ones that handle time series data because of the nature of brain waves. However, many of these time series models, like HMM with hidden state in discrete space or State Space Model with hidden state in continuous space, only work with one source of data and cannot handle different sources of information simultaneously. In this paper, we propose an extension of State Space Model to work with different sources of information together with its learning and inference algorithms. We apply this model to decode the mind state of students during lectures based on their brain waves and reach a significant better results compared to traditional methods. 

\end{abstract}

\section{Introduction}
\label{sec:intro}

Understanding how human brain functions has been an attractive research question in recent years \cite{mitchell2008predicting}. One important progress is on collecting a large amount of brain wave data with different technologies, like fMRI \cite{wehbe2014simultaneously}, MEG \cite{sudre2012tracking} and EEG \cite{wang2013using}. The nature of these data collecting technologies have introduced a variety of substantial challenges in understanding these data with machine learning techniques. For example, MEG and EEG technologies can describe the brain with considerable temporal granularity, but with a relatively low spatial resolution. Therefore, machine learning techniques that can handle temporal dependencies are highly appreciated. 

Fortunately, in recent years, there is an increasing trend towards the use models to work with time series problem. 
For example, \cite{khaleghi2013nonparametric} find the points in time where the probability distribution generating the data has changed given a heterogeneous time-series sample. 
\cite{anava2013online} use regret minimization techniques to develop effective online learning algorithms for predicting a time series using autoregressive moving average model.
\cite{alon2003discovering} fit a finite mixture of HMMs in motion data, using the expectation maximization (EM) framework, aiming at discovering groupings of similar object motions that were observed in a video collection.
Also, Hidden Markov Models or other graph-based methods are often used for the purpose like speech recognition, pattern recognition and neural networks, similar to \cite{nahar2016arabic}, \cite{schwenk1999using}, \cite{bar2004analyzing} and \cite{lecun1995convolutional}.
In a more rigorous setting, a lot more theoretical questions are being asked and solved for time series machine learning these days. For example, \cite{khaleghi2013nonparametric,khaleghi2014asymptotically} derive theories for estimation of highly dependent time series data.
\cite{kuznetsov2014generalization,kuznetsov2015learning} push the theoretical work further for non-stationary time series problems.  
\cite{rakhlin2013online,anava2013onlinekuznetsov2016time} naturally combines the problem of analyzing time series data with online learning, which opens the door to a whole area of new problems. 

With the guidance of previous work on time series data. In this paper, we present Sequence Data based Mind-Detecting (SeDMiD) Model 
, a novel time series method to uncover the state of brain using more than a typical source of EEG/MEG recordxings. Others sources like video recordings or audio data can be supplemented for better performance on brain state estimation.

We also develop the learning algorithm for SeDMiD based on sparsity regularized linear system, and the inference algorithm as an extension of Viterbi algorithm under Gaussian assumption for continuous space. 
The results show that, the performance of SeDMiD can uncover the mind state based on brain waves with a significant better results than traditional methods. 

Our contribution of this paper is three-fold:
\begin{itemize}
\item We propose a SeDMiD that can analyze time series brain wave data with extra sources of information. 
\item We improve the existing vertibi algorithm to enable the inference of SeDMiD model.
\item We show the possibility of deciphering students' mind state of understanding lectures with brain wave data. 
\end{itemize}

The rest of this paper is organized as following. Section~\ref{sec:relatedwork} describes some work that others did to solve the prediction problem using MEG data. Section~\ref{sec:method} raises the novel model and describe its learning and inference method, while section~\ref{sec:exp} shows the result of our experiment. Finally, Section~\ref{sec:con} concludes and suggests future work.

\section{Related Work}
\label{sec:relatedwork}
There are many works implementing time series technology to deal with the problem about MEG/EEG data. It is worth noting that both electroencephalography (EEG) and magnetoencephalography (MEG) provide a more direct measure of the electrical activity in the brain, as is described professionally in the work of \cite{michel2009electrical} and \cite{proudfoot2014magnetoencephalography}. \cite{he2008electrophysiological} measure the difference in electric potentials on the scalp and captures high frequency oscillations on the millisecond timescale that is most relevant for the characterisation of cognitive processes. There are many existing work to do experiments and discover new knowledge in several fields.
\cite{moghadamfalahi2015language} use abstract—noninvasive EEG-based brain–computer interfaces (BCI) for intent detection, specifically for EEG-based BCI typing systems.
\cite{phillips1997imaging} have developed a Bayesian framework for image estimation from combined MEG/EEG data.

EEG signal is a kind of voltage signal that can be measured on the surface of the scalp, arising from large areas of coordinated neural activity manifested as synchronization (groups of neurons firing at the same rate), described by \cite{niedermeyer2005electroencephalography}. 
This neural activity varies as a function of development, mental state, and cognitive activity, and the EEG signal can measurably detect such variation
\cite{marosi2002narrow}, \cite{lutsyuk2006correlation}, \cite{berka2007eeg}, \cite{wang2013using} which in turn are important for and predictive of learning \cite{baker2010better}. 

On the other hand, hidden Markov model is also widely implemented to make best use of MEG/EEG recordings.
\cite{rukat2016resting} analyses the temporal and spatial dynamics of physiological substrate of cognitive processes as measured by EEG, with a hidden Markov model. 
\cite{liu2010eeg} combine kernel principal component analysis (KPCA) and HMM to differentiate mental fatigue states with the help of EEG data.
\cite{ko2011hsa} describe a procedure of classification of motor imagery EEG signals using HMM, which can tell the person is performing left, right hand or foots motor imagery based on the current EEG recordings.
What is more, another bio-signal named electrocardiogram (ECG) is also suitable for Markov model.
\cite{coast1990approach} and \cite{andreao2006ecg} describe a new approach to ECG arrhythmia analysis based on HMM.
Inspired by these work, we propose a novel model to analyze time series brain wave data with extra sources of information. 

\section{SeDMiD Model and Its Learning and Inference Algorithm}
\label{sec:method}

We propose a novel state-space model for inferring people's state using several sources of information simultaneously including MEG/EEG recordings, named Sequence Data based Mind-Detecting (SeDMiD) Model. In this section, we will first introduce the model, and then show its learning and inference algorithm respectively.

\subsection{SeDMiD}
The aim of our work is to raise a model for brain-state estimation using MEG/EEG recordings while considering other essential time-series sources at the same time.
A pictorial representation of our model can be found in Figure 1. Notations of this paper is illustrated in the caption of Figure 1.

In the SeDMiD model, we firstly assume that the extra time-series sources can match exactly with brain wave signals in aspect of time stamps, which means $\vec{S_i}$, $\vec{C_i}$ and $\vec{M_i}$ are sampled at the same time. And we simplify the mind state inferring task as a binary decision of states $\vec{C_i}$, indicating that people is either happy or sad, or students are either confused or not.
We also assume that MEG/EEG recordings $\vec{M_i}$ have linear dependence on current external source and current mind state, since brain signal is easily affected by people's internal mind situation and outside influence on them \cite{baker2010better}.
Additionally, current mind state $\vec{M_i}$ depends linearly on previous status and current video contents, because brain state often has a close connection with its state few seconds ago, and other sources like videos also affect the mind state. 
Finally, the complementary source is a continuous time-series process, thus current recording is dependent on the former one, and to simplify the model, we assume linear dependency again.

\begin{figure}
\label{figure1}
\caption{The proposed model. For the time-stamp at $i$, $\vec{M_i}$ corresponds to EEG/MEG recording and $\vec{C_i}$ corresponds to mind states (e.g. happy/sad, confused/non-confused). $\vec{S_i}$ is extra source information (e.g. video/audio data). $\vec{A}$, $\vec{B}$, $\vec{E}$, $\vec{F}$ and $\vec{G}$ are stationary transition matrices that describe the dynamics among states in the graph.}
\centering
    \includegraphics[width=12cm]{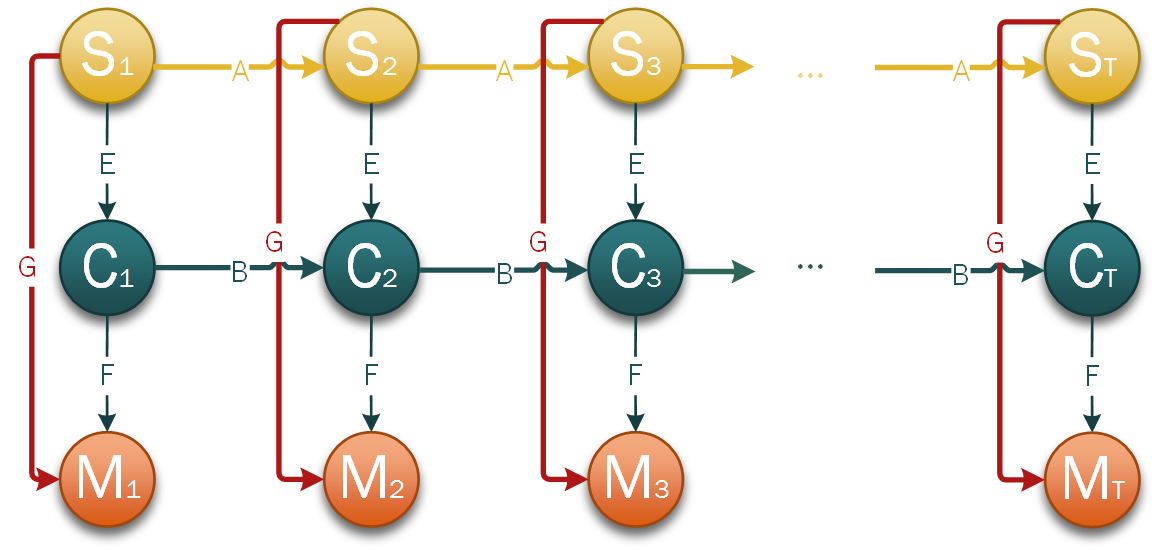}
\end{figure}

Further, we assume that the supplemental source data $\{\vec{S_i}\}_{i=1}^n$ can be described by Gaussian distribution.
\begin{equation*}
    \vec{S_1} \thicksim \mathcal{N}( \vec{\mu_s} , \vec{\sigma_s} )\\
\end{equation*}
\begin{equation*}
  \vec{S_i} \thicksim \mathcal{N}( \vec{A} \cdot \vec{S_{i-1}}, ~\vec{\Lambda_i} )
\end{equation*}

We also assume that mind states $\{\vec{C_i}\}_{i=1}^n$ is in form of Gaussian distribution.
\begin{equation*}
    \vec{C_1} \thicksim \mathcal{N}( \vec{\mu_c} , \vec{\sigma_c} )
\end{equation*}
\begin{equation*}
    \vec{C_i} \thicksim \mathcal{N} \left\lgroup \bm{l(} \vec{E} \cdot \vec{S_i} + \vec{B} \cdot \vec{C_{i-1}} \bm{)},  ~\vec{\Omega_i} \right\rgroup 
\end{equation*}
Considering the assumption that only on-and-off state exist in space $\set{C}$, we use the function $\bm{l()}$ to map the score of positive state ($ \vec{E} \cdot \vec{S_i} + \vec{B} \cdot \vec{C_{i-1}}$) into two dimensions above.
\begin{equation*}
  \bm{l(}x\bm{)} = 
    \begin{bmatrix}
        \frac{1}{1+e^x}     \\
        \frac{e^x}{1+e^x}
    \end{bmatrix}
\end{equation*}
Thus brain wave recordings $\{\vec{M_i}\}_{i=1}^n$ also follows Gaussian distribution because it is the linear combination of $\{\vec{S_i}\}_{i=1}^n$ and $\{\vec{C_i}\}_{i=1}^n$, as the mean of $\vec{M_i}$ is $\vec{G} \cdot \vec{S_i} + \vec{F} \cdot \vec{C_i}$ and co-variance is $\vec{\Sigma_i}$.
\begin{equation*}
    \vec{M_i} \thicksim \mathcal{N}( \vec{G} \cdot \vec{S_i} + \vec{F} \cdot \vec{C_i}, ~\vec{\Sigma_i} ) 
\end{equation*}
The practical meaning of notations in the model is described by Table 1.
\begin{table}[hbtp]
\label{table:t1}
\floatconts
  {MEG Feature}
  {\caption{Practical Meaning of Notations in SeDMid model}}
  {\begin{tabular}{cl}
  \toprule
  \bfseries Notation & \bfseries Description\\
  \midrule
  $ \vec{S_i} $   & Features of supplemental source information at time $i$ \\
  $ \vec{C_i} $   & Mind States at time $i$       \\
  $ \vec{M_i} $   & Brain wave signal recordings (EEG/MEG) at time $i$    \\
  $ \vec{A} $   & Linear Relationship between adjacent supplemental source information \\
  $ \vec{B} $   & Linear Relationship between adjacent mind states  \\
  $ \vec{E} $   & Linear Relationship between supplemental source and mind state \\
  $ \vec{F} $   & Linear Relationship between mind state and brain wave signal  \\
  $ \vec{G} $   & Linear Relationship between supplemental source and brain wave signal   \\
    \bottomrule
  \end{tabular}}
\end{table}

\subsection{Learning}
Here we introduce the parameter learning algorithm of $\vec{A}$, $\vec{B}$, $\vec{E}$, $\vec{F}$and $\vec{G}$ for SeDMiD. Because of the nature of the task, a supervised training training procedure, with $\vec{S}$, $\vec{C}$, $\vec{M}$ known, is sufficient. 
At training, at time-stamp $i$, we can observe MEG/EEG recording $\vec{M_i}$, supplemental source data $\vec{S_i}$ and current mind state $\vec{C_i}$.
Thus, parameter learning is a maximum likelihood estimation (MLE) problem, in which case linear regression is the solution. 
However, supplemental source data generally comes with a higher dimension than response variable space. 
Therefore, sparsity regularize is required for transition matrix $\vec{S}$ by \equationref{eq:learning_A} in a sparse form. 
\begin{equation}
\label{eq:learning_GF}
    \vec{M_i} =
    \begin{bmatrix}
        \vec{G} & \vec{F}
    \end{bmatrix}
    \begin{bmatrix}
        \vec{S_i} \\
        \vec{C_i}
    \end{bmatrix}
\end{equation}

\begin{equation}
\label{eq:learning_BE}
    \vec{C_i} = 
    \begin{bmatrix}
        \vec{B} & \vec{E}
    \end{bmatrix}
    \begin{bmatrix}
        \vec{C_i} \\
        \vec{S_{i+1}}
    \end{bmatrix}
\end{equation}

\begin{equation}
\label{eq:learning_A}
    \vec{S_{i+1}} = \vec{A} \cdot \vec{S_i}
\end{equation}

\subsection{Inference}
In inference, observation only contains brain wave recordings, so we want to estimate mind state $\vec{C_i}$ and even further inference complementary data $\vec{S_i}$. With the observation $\vec{M_i}$ and five transition matrices that learned in the learning phase, we formulate an inference method to find the best sequence of mind states $\vec{C_1}, \vec{C_2}, \dots, \vec{C_T}$ and features of complementary data $\vec{S_1}, \vec{S_2}, \dots, \vec{S_T}$. A natural choice for inference on the model is Viterbi algorithm \cite{forney1973viterbi}. However, the problem is that the state space $\set{S}$ and $\set{M}$ are infinite, and we assume that $S_i$ and $M_i$ follow Gaussian distribution, we formulate the inference mathematically in Gaussian form. Consider the Viterbi function at the time-stamp $t$ is calculated by \equationref{eq:viterbi_1}, given by the factorization of graphic model shown in \figureref{figure1}.
\begin{equation}
\label{eq:viterbi_1}
    \vec{V_t}(\vec{S_t} , \vec{C_t}) = \max_{\vec{S_{t-1}},\vec{C_{t-1}}}[P(\vec{M_t}|\vec{S_t}, \vec{C_t})P(\vec{S_t}|\vec{S_{t-1}})P(\vec{C_t}|\vec{S_t},\vec{C_{t-1}})V_{t-1}(\vec{S_{t-1}},\vec{C_{t-1}})]
\end{equation}
To begin with, we consider to calculate the Viterbi function $\vec{V_1}$ at the start point of the whole process when t = 1.
\begin{equation}
    \begin{split}
        \vec{V_1}(\vec{S_1}, \vec{C_1}) &= P(\vec{M_1}|\vec{S_1},\vec{C_1})P(\vec{S_1})P(\vec{C_1}|\vec{S_1})\\
                      &= \mathcal{N}(\vec{G} \cdot \vec{S_1} + \vec{F} \cdot \vec{C_1},  \vec{\Sigma}) \cdot \mathcal{N}(\vec{\mu_s}, \vec{\sigma_s}) \cdot \mathcal{N}(\bm{l(} \vec{E} \cdot \vec{\mu_s}\bm{)}, \vec{\sigma_c} ) 
    \end{split}
\end{equation}

According to the property of multiplication in Gaussian distribution, we obtain that $\vec{V_1}$ is also in form of Gaussian with its mean and variance shown in \eqref{eq:v_1} and \eqref{eq:V_1} respectively.

\begin{equation}
  \label{eq:v_1}
    \vec{\Sigma_{V_1}} = \left\lgroup
        \begin{bmatrix}
        \vec{\sigma_s}^{-1} & \vec{0} \\
        \vec{0} & \vec{\sigma_c}^{-1}
        \end{bmatrix}
    +
        \begin{bmatrix}
        \vec{G}^T     \\
        \vec{F}^T
        \end{bmatrix}
        \vec{\Sigma}^{-1}
        \begin{bmatrix}
        \vec{G} & \vec{F}
        \end{bmatrix}
        \right
        \rgroup
    ^{-1}
\end{equation}

\begin{equation}
  \label{eq:V_1}
    \vec{\mu_{V_1}} = 
    \vec{\Sigma_{V_1}}\left\lgroup
    \begin{bmatrix}
    \vec{\sigma_s}^{-1}\vec{\mu_s}\\
    \vec{\sigma_c}^{-1}\bm{l(}\vec{E} \cdot \vec{\mu_s}\bm{)}
    \end{bmatrix}
    +
    \begin{bmatrix}
    \vec{G}^T\\
    \vec{F}^T
    \end{bmatrix}
    \vec{\Sigma}^{-1}
    \vec{M_1}
    \right\rgroup
\end{equation}

For now, the mean and covariance of Viterbi function for the beginning point have been calculated, and in order to know all the Viterbi function at every time-stamp, we find when $t \geqslant 2$:
\begin{equation}
    \vec{V_t}(\vec{S_t}, \vec{C_t}) = \max_{\vec{S_{t-1}},\vec{C_{t-1}}}P(\vec{C_t}|\vec{S_t},\vec{C_{t-1}})P(\vec{S_t}|\vec{S_{t-1}})V_{t-1}(\vec{S_{t-1}},\vec{C_{t-1}})P(\vec{M_t}|\vec{S_t},\vec{C_t})
\end{equation}
Also, we can calculate the mean and covariance since for time $t$, $\vec{S_t}$ and $\vec{C_t}$ are in the form of Gaussian distribution.
\begin{equation}
\label{eq:v_2}
    \vec{\Sigma}_{\vec{V_2}} = 
        \left\lgroup
        \vec{\Phi^{-1}}
        +
        \begin{bmatrix}
        \vec{G}^T     \\
        \vec{F}^T
        \end{bmatrix}
        \vec{\Sigma}^{-1}
        \begin{bmatrix}
        \vec{G}  &  \vec{F}
        \end{bmatrix}
        \right\rgroup^{-1}
\end{equation}

\begin{equation}
    \vec{\mu}_{\vec{V_2}} = 
        \vec{\Sigma}_{V_2}\left\lgroup
        \vec{\Phi^{-1}} \cdot \vec{\mu}_{\vec{V_1}}
        +
        \begin{bmatrix}
        \vec{G}^T     \\
        \vec{F}^T
        \end{bmatrix}
        \vec{\Sigma}^{-1}
        \vec{M_2}
        \right\rgroup
\end{equation}

where

\begin{equation}
\label{eq:phi}
\vec{\Phi} = \left\lgroup
\begin{bmatrix}
\vec{\Lambda}^{-1} & \vec{0} \\
\vec{0} & \vec{\Omega}^{-1}
\end{bmatrix}
+
\left\lgroup
\begin{bmatrix}
(\vec{A}^{-1})^T & \vec{0}    \\
\vec{0} & (\vec{B}^{-1})^T
\end{bmatrix}
\vec{\Sigma}_{V_1}^{-1}
\begin{bmatrix}
\vec{A}^{-1} & \vec{0}  \\
\vec{0} & \vec{B}^{-1}
\end{bmatrix}
\right\rgroup
\right\rgroup^{-1}
\end{equation}

Once the means and co-variances are calculated for all supplemental source data and mind state sequence, the best sequence of $\vec{S}$ and $\vec{C}$ can be inferred by calculating backwards. So we start the inference at the end of the time sequence.
\begin{equation}
    \begin{bmatrix}
    \vec{S_T} \\
    \vec{C_T}
    \end{bmatrix}
    = \arg \max_{\vec{S_T},\vec{C_T}}\mathcal{N}(\vec{\mu}_{V_T}, \vec{\Sigma}_{V_T})
\end{equation}
Note that the mean of the final state is calculated exactly, we proceed to infer the former state that lead to the current one, so the formula to calculate the former state based on current state is shown in \eqref{eq:t-1}.
\begin{equation}
\label{eq:t-1}
    \begin{bmatrix}
    \vec{S_{T-1}} \\
    \vec{C_{T-1}}
    \end{bmatrix}
    = \arg \max_{\vec{S_{T-1}},\vec{C_{T-1}}}\mathcal{N}(\vec{\gamma}, \vec{\tau})
\end{equation}
where
\begin{equation}
    \vec{\tau} = \left\lgroup
        \vec{\Sigma}_{V_{T-1}}^{-1}
        +
        \begin{bmatrix}
        (\vec{E} \cdot \vec{A})^T \\
        \vec{B} ^ T
        \end{bmatrix}
        \vec{\Omega}^{-1}
        \begin{bmatrix}
        \vec{E} \cdot \vec{A} & \vec{B}
        \end{bmatrix}
        +
        \begin{bmatrix}
        \vec{A}^T\vec{\Lambda}^{-1}\vec{A} & \vec{0}    \\
        \vec{0} & \vec{0}
        \end{bmatrix}
        \right\rgroup^{-1}
\end{equation}

\begin{equation}
    \vec{\gamma} = \vec{\tau}\left\lgroup
        \vec{\Sigma}_{V_{T-1}}^{-1}\vec{\mu}_{V_{T-1}}
        +
        \begin{bmatrix}
        (\vec{E} \cdot \vec{A})^T \\
        \vec{B} ^ T
        \end{bmatrix}
        \vec{\Omega}^{-1}
        \begin{bmatrix}
        \vec{S_T} \\
        \vec{C_T}
        \end{bmatrix}
        +
        \begin{bmatrix}
        \vec{A}^T\vec{\Lambda}^{-1}\vec{A} & \vec{0}    \\
        \vec{0} & \vec{0}
        \end{bmatrix}
        \begin{bmatrix}
        \vec{S_T} \\
        \vec{C_T}
        \end{bmatrix}
        \right\rgroup^{-1}
\end{equation}
Finally, by calculate all the $\vec{S_i}$ and $\vec{C_i}$ from end to the first one, we can infer mind states for each time-stamp $t \in \{1, \dots, T\}$.

\section{Experiment}
\label{sec:exp}
To solve the problem that we raised in \sectionref{sec:problemDef}, we need to extract features of lecture videos. Using SeDMid model that we raise in \sectionref{sec:method}, 

\subsection{Experiment Setting}
\label{sec:problemDef}
In our experiment, we set up a task to estimate students' mind states in a given lecture session, finding out they are confused or not with the data from \cite{wang2013using}. Every participants is asked to watch 10 lecture videos, the length of which is around 2 minutes.The two-minute period is called an 'experiment period'. In all there are 10 students participate in the experiments so there are 100 data points in total. 
During every experiment period, their EEG signals are recorded in the frequency of 1 Hz, and EEG signals here have 11 features, including Proprietary measure of mental focus, 1-3 Hz of power spectrum and so on.
Students are asked to annotate whether they are confused or not based on every whole experiment period, and every second in the period by the annotation is noted.
The beginning and the end of every experiment is cut off with consideration of reducing noise, only left the middle 112 seconds for analysis.
Finally, lecture videos are served as supplemental source data.

\subsection{Video Feature Extraction}

We use the tool kit OpenCV developed by \cite{bradski2008learning} to extract video features like optical features and object movement information, and openSMILE developed by \cite{eyben2010opensmile} to extract features in audio data, such as lecturer's speech speed and intonation. As a result, we obtain video image features with 1440 dimensions while audio features have 6669 dimensions, thus we get 8109 dimensions for video features in total. Since both MEG recordings and students' status vector is collected in the frequency of 1 Hz, we sample the feature vector every 1 second in order to alignment the data for SeDMiD model. 

\subsection{Performance Comparison}
We use the simple state space model (SSM) as baseline, which only make use of EEG data, and current mind state only depends on the former state, while current EEG recording depends on current mind state. SSM does not make use of video features. We also compare our model with logistic regression, which employ brain wave signals without other sources.
As is shown in the ROC curve in Figure 2, we find that our SedMid model outperform the simple HMM model and logistic regression, with the accuracy of ours reaches 87.76\% while simple HMM only gets 53\% and logistic regression 60\%.
In the experiment, We find that errors that SeDMiD makes always exist at the beginning of experiment period, and it will lead to correct result in few time. It is intuitive that SeDMiD model often makes mistakes at beginning of every experiment period because it can perform better with more data gets concluded. 
\begin{figure}[h]
\label{fig:1}
\caption{The ROC curve for comparing the performance of SeDMiD model, Simple State Model (SSM) and logistic regression. SeDMiD greatly outperforms the others.}
\centering
    \includegraphics[width=12cm]{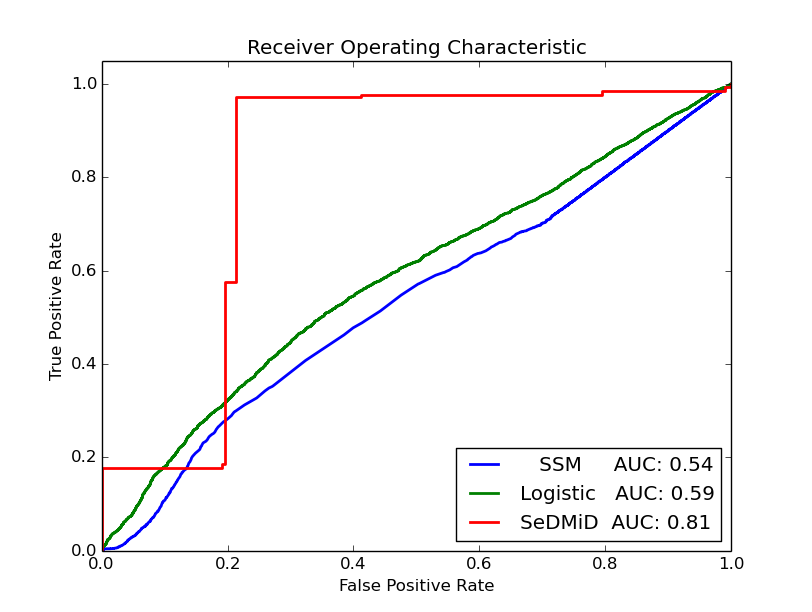}
\end{figure}


\section{Conclusion}
\label{sec:con}
In this work, we propose an novel state space model called Sequence Data based Mind-Detecting (SeDMiD) Model, which analyzes time series brain wave data with extra sources of information, improving the existing vertibi algorithm to enable the inference. We evaluate the effectiveness of SeDMiD model by comparing with simple Markov model and logistic regression. The performance of our model has a 30\% higher in accuracy than the normal one. 

Apart from proposing the SeDMiD model, our contribution includes showing the possibility of deciphering students' mind state of understanding lectures with brain wave data. 
This work is also useful in real world implementation, since teachers can modify their teaching strategy based on audiences' status if it can be inferred.

\acks{This work was supported by International School, Beijing University of Posts and Telecommunications. We thank Hao Ding for advising the method of extracting videos' features and dormitory administrator in Student Building 5, BUPT, for providing electronic power for running experiment at night.}

\bibliography{main}



\end{document}